\documentclass[preprint,nofootinbib]{revtex4}%
\usepackage{amssymb}
\usepackage{amsfonts}
\usepackage{amsmath}
\usepackage{amsmath}
\usepackage{graphicx}
\usepackage[usenames]{color}%
\usepackage[titletoc,toc,title]{appendix}
\providecommand{\U}[1]{\protect\rule{.1in}{.1in}}
\providecommand{\U}[1]{\protect\rule{.1in}{.1in}}
\definecolor{blue}{rgb}{0,0,1}

\definecolor{red}{rgb}{1,0,0}

\usepackage{hyperref}
\hypersetup{colorlinks=true,allcolors=blue,	pdftitle={Critical magnetic field in the strong regime for 3+1 holographic superconductors}}

\begin{document}

\title{Critical magnetic field in the strong regime for 3+1 holographic superconductors}
\author{Mariano Chernicoff$^1$, Cesar D\'iaz$^1$, Leonardo Pati\~no$^1$ and Mart\'in Reyes$^1$}
\affiliation{$^1$Departamento de F\'{\i}sica, Facultad de Ciencias, Universidad Nacional Aut\'{o}noma de M\'{e}xico,   A.P. 70-542, CDMX 04510, M\'{e}xico.}

\begin{abstract}

In this letter, we explore different aspects of a $3+1$ strongly coupled superconductor in the presence of an intense magnetic field. In order to study this interesting and highly complicated system, we make use of the holographic correspondence. We construct a gravitational solution incorporating a fully backreacted Maxwell field while treating a charged scalar field as a perturbation propagating on this background. Our work focuses on the implications of intense magnetic fields on superconductivity in the strongly coupled regime, offering insights into the modifications that such a field induces on key aspects of the superconducting phase. Through rigorous numerical analysis of the holographic model, we uncover intriguing effects arising from the interplay between the magnetic field and the superconducting condensate.

\end{abstract}
\maketitle

\section{Introduction}
\label{Intro}

Superconductivity, the remarkable quantum phenomenon characterized by the complete loss of electrical resistance and the expulsion of magnetic fields, has fascinated physicists for decades due to its fundamental importance and technological potential (e.g. \cite{tinkham2004introduction, poole2014superconductivity}). Despite its widespread applications, understanding the microscopic mechanisms underlying superconductivity remains a formidable challenge, especially in materials exhibiting strong electronic correlations and high critical temperatures. The complexity of strongly coupled quantum field theories governing these systems often defies conventional analytical methods, motivating the exploration of alternative theoretical frameworks such as the holographic correspondence \cite{Maldacena:1997re}.

The difficulty in modeling and understanding superconductivity arises from the intricate interplay of quantum mechanical effects, including electron pairing mediated by lattice vibrations (phonons) or other collective excitations. Conventional approaches, such as BCS theory \cite{Bardeen:1957mv}, provide a phenomenological description but fall short in capturing the behavior of materials where strong interactions dominate. In such scenarios, the holographic correspondence offers a promising avenue for unraveling some of the mysteries behind strongly coupled quantum matter and in particular, it provides a unique lens to study the phenomenon of superconductivity.

The superconducting state exists below a system-dependent critical temperature $T_c$. As we already mentioned, it can be characterized by the complete disappearance of electrical resistance and the expulsion of magnetic fields from the interior of the material that has reached such a state. Loosely speaking, from the quantum field theory point of view, the transition to the superconducting phase can be understood as the spontaneous breaking of a gauge symmetry which is signalized by the condensation of a generically composite, charged operator $\mathcal{O}$ at low temperatures $T < T_c$ \cite{Nambu:1960tm, Anderson:1963pc}.
As we will explain below, when considering more general scenarios, the superconductive phase can also be affected by some other physical parameter of the theory like the presence of a magnetic field that typically leads to the suppression of superconductivity in conventional systems \cite{Ginzburg:1950sr, abrikosov1957magnetic}.

Significant progress has been made in studying holographic superconductors in three spacetime dimensions (see for example \cite{Cai:2015cya, Hartnoll:2020fhc}), however, a more realistic approach involves considering materials in four-dimensional spacetimes that mirror our physical universe. This extension provides deeper insights into the behavior of real-world superconductors. Recent studies have begun exploring the effects of strong magnetic fields on holographic superconductors in four dimensions, revealing intriguing phenomena such as chemical potential-induced phase transitions and modifications in critical temperatures, e.g. \cite{albash2008holographic, nakano2008critical, dector2015magnetic,Correa:2019ivh}. 

Deeply motivated by these ideas, in this letter, we have constructed a $3+1$ holographic superconductor and studied its behavior in the presence of an arbitrarily intense magnetic field.

The paper is structured as follows: Section \ref{secbuilding} provides a brief overview of the key elements required to construct a holographic model dual to a superconductor. In Section \ref{secbulkequations} we examine the equations and symmetries of the model, introducing the consistent ansatz used in our numerical analysis. Section \ref{secnumresults} details the numerical scheme employed, which is then applied to explore the properties of both the normal and superconducting phases using the holographic framework. Section \ref{secresults} concludes with a summary of our results.

\section{Building a Holographic Superconductor}
\label{secbuilding}

In the context of the holographic correspondence, the simplest gravitational setup dual to a field theory which undergoes a phase transition that presents features of superconductivity was first presented in \cite{Gubser:2008px, Gubser:2008zu}. The basic ingredients needed to build a 4-dimensional holographic superconductor are listed in these beautiful papers, and we implemented them in the following manner. The general structure is that of an asymptotically $AdS_5$ black hole in which the Hawking temperature $T_h$ is dual to the corresponding temperature of the 4-dimensional CFT. We also need to include a gauge field $A_{\mu}$ and a massive charged scalar field $\psi$ in the bulk, which according to the AdS/CFT correspondence are dual to a current $J^{a}$, and a gauge invariant operator $\mathcal{O}$ at the boundary, respectively. 

The elements above can be consistently incorporated through the gravity action \cite{Hartnoll:2008vx, Hartnoll:2008kx, Hartnoll:2009sz} given by
\begin{equation}
S = \frac{1}{16 \pi G_5} \int d^{5}x \sqrt{-g} \left( R - \Lambda - \frac{1}{4}F_{\mu \nu}F^{\mu \nu}
 -\left\vert \nabla \psi - i \xi A \psi \right\vert^2 - m^2 |\psi|^2 \right) + \frac{k}{12\pi G_5} \int A\wedge F \wedge F \ ,
\label{gubseraction}
\end{equation}
where $\Lambda=12/L^2$, $L$ is the radius of curvature of $AdS_5$, while $m$ and $\xi$ are the mass and charge of the complex scalar field respectively. The field strength is defined by $F_{\mu \nu}=\partial_{\mu}A_{\nu}-\partial_{\nu}A_{\mu}$, with $\mu,\nu=0,\ldots , 4$. The last term in (\ref{gubseraction}) corresponds to the so-called Chern-Simons term where the parameter $k$ is known as the level of the theory \cite{Deser:1981wh, Deser:1982vy}.

As mentioned in the Introduction, in this letter, we are interested in building a holographic superconductor in the presence of a strong magnetic field. Therefore, we need a solution to the equations following from \eqref{gubseraction} that fully includes the backreaction of the gauge field $A_{\mu}$. The presence of a complex scalar field in (\ref{gubseraction}) is also crucial because, from the field theory point of view, it allows the formation of a charged condensate with a non-trivial phase structure. For reasons that will be clarified below, it is consistent to neglect the backreaction of the complex scalar field and simply solve the corresponding equation of motion on such fixed background. This perturbative approach simplifies our analysis by reducing the computational complexity, focusing on the essential physics of the gauge and gravitational fields while still capturing the critical effects of superconductivity. It is important to point out that our presentation below suffices only to complete such perturbative approach for the scalar field, and as we will see, in this sense it also limits the thermodynamic region we are able to explore.

\section{Bulk equations for a holographic superconductor}\label{secbulkequations}
 
To tackle the challenge of solving the aforementioned equations of motion, we will employ numerical methods that are essential not only for managing the intricate equations that arise from the action but also for exploring the parameter space where interesting physical phenomena might occur, e.g. the superconducting phase. 

Starting from (\ref{gubseraction}) and fixing $L=1$ and $k=1$, the corresponding equations of motion are given by, 
\begin{align}
R_{\mu \nu} & = \frac{1}{2} R g_{\mu \nu} - 6 g_{\mu \nu} + \frac{1}{2} F_{\alpha \beta} F^{\alpha \beta} g_{\mu \nu} + \frac{1}{2} |\partial \Psi - i \xi A \Psi |^2 g_{\mu \nu} + \frac{1}{2} m^2 |\Psi|^2 g_{\mu \nu} - 2 F_{\mu}^{a} F_{\nu a}\nonumber \\
 & \qquad {} - \frac{1}{2}\left( \partial_{\mu} \Psi - i \xi A_{\mu} \Psi \right) \left( \partial_{\nu} \bar{\Psi} + i \xi A_{\nu} \bar{\Psi}\right) - \frac{1}{2}\left( \partial_{\nu} \Psi - i \xi A_{\nu} \Psi \right) \left( \partial_{\mu} \bar{\Psi} + i \xi A_{\mu} \bar{\Psi}\right), \label{einstein} \\
\nonumber \\
0  & =  i \xi A_{\mu} (\partial_{\nu} \Psi - i \xi A_{\nu} \Psi )g^{\mu \nu} + m^2 \Psi   - \frac{\partial_{\mu} \left[ \sqrt{-g} g^{\mu \nu} ( \partial_{\nu} \Psi - i \xi A_{\nu} \Psi ) \right]}{\sqrt{-g}},  \label{psi} \\
\nonumber \\
0 & = 4 \partial_{\nu} \left( \sqrt{-g}F^{\nu \mu} \right) + 2 \epsilon^{\alpha \beta \gamma \delta \mu} F_{\alpha \beta} F_{\gamma \delta} \nonumber \\
 & \qquad {} - \sqrt{-g} g^{\mu \nu} \left[ i \xi \left( \overline{\Psi} \partial_{\nu} \Psi - \Psi \partial_{\nu} \overline{\Psi} \right) + 2 \xi^{2} A_{\nu} |\Psi|^2 \right]
 \label{fulleom}. 
\end{align}
The equation of motion for $\bar{\Psi}$ is simply the complex conjugate of equation (\ref{psi}).

For the sake of clarity, it is worth emphasizing that \eqref{einstein} to \eqref{fulleom} describe the fully backreacted system, however, according to what we briefly explained in the previous section, in order to study certain aspects of a holographic superconductor in the presence of a strong magnetic field, it suffices to consider the complex scalar field only as a perturbation propagating on a fixed background. Taking this into account, in what follows, and before the computational numerical methods kick in, we will propose an ansatz for the metric $g_{\mu\nu}$ and the gauge field $A_{\mu}$ that is consistent with the symmetries of our construction and will greatly simplify the final solution. 

Preparing to accommodate the necessary field content, the first thing to require is for the metric to be invariant under the following global symmetries: i) Translation invariance in the $x, y, z$ coordinates, ii) rotational invariance in the $z$ coordinate and iii) boost invariance in the $z$ direction.

Also, as customary in the context of the holographic correspondence, our ansatz needs to satisfy the standard Brown-Henneaux asymptotic boundary conditions \cite{Brown:1986nw}. 
Under these considerations, we can write an ansatz for the metric given by

\begin{equation}
ds^2 = -U(r) dt^2 + \frac{dr^2}{U(r)} + V(r) ( dx^2 + dy^2 ) + W(r) (dz + C(r) dt)^2,
\label{metricansatz}
\end{equation}
where $0\leq r \leq \infty$ is the holographic radial coordinate, and $t, x, y, z$ correspond to the time and spatial coordinates in the dual CFT.
Given the reparametrization invariance under the $r$ coordinate, we can set the metric functions $g_{tt} = -g_{rr}^{-1} = U(r)$. It is also worth mentioning that any rescaling of the $x$ and $y$ coordinates can be compensated by rescaling the metric function $V(r)$. Equivalently, any rescaling of the $z$ coordinate can be compensated by rescaling $C(r)$ and $W(r)$.

Let us now focus on the gauge field $A_{\mu}$. Notice that given (\ref{metricansatz}), and considering the global symmetries mentioned above, the ansatz for the field strength $F_{\mu\nu}$ can be written as
\begin{equation}
F= E(r) dr \wedge dt + b dx \wedge dy + P(r) dz \wedge dr,
\label{Fansatz}
\end{equation}
which in a suitable gauge is obtained from the vector potential
\begin{equation}
A = A_t dt + A_y dy +A_z dz ,
\end{equation}
where
\begin{equation}
\begin{split}
A_t  & = \int E(r) dr, \\
A_y  &= b x, \\
A_z  &= \int P(r) dr.
\label{Aa}
\end{split}
\end{equation}
The $y$ component in (\ref{Aa}) is linked to a constant magnetic field $b$ in the bulk. Less obvious is the role played by the functions $E(r)$ and $P(r)$, but as we will explain later in the text, these are jointly related to the electric charge density and the chemical potential in the dual CFT. From (\ref{Aa}), it follows that the equations of motion \eqref{einstein} to \eqref{fulleom} are invariant (on-shell) under the local transformations,
\begin{equation}
\begin{split}
z & \to z- \alpha t, \\
C(r) & \to C(r) + \alpha, \\
E(r) & \to E(r) - \alpha P(r),
\label{localsymmetry}
\end{split}
\end{equation}
with $\alpha \in \mathbb{R}$. This type of local symmetry is known as $\alpha$-symmetry \cite{DHoker:2009ixq, DHoker:2009mmn}. The important role played by (\ref{localsymmetry}) will be better understood when the numerical computations begin, however, as a first glimpse let us notice that the value of $\alpha$ can be chosen so that $C(r\rightarrow \infty)=0$, which can be used to take (\ref{metricansatz}) to an asymptotically  $AdS_5$ form. Also, the function $\mathcal{E} = E+CP$ is invariant under $\alpha$-symmetry. Therefore, it will be convenient to express some equations in terms of this function.

As mentioned in the Introduction, testing the superconducting phase in the context of holography requires examining the dynamics of a complex scalar field as part of certain gravitational configurations. Given our particular setup, it can be shown that it is consistent with (\ref{fulleom}) to study the dynamics of the scalar field $\Psi$ perturbatively to characterize some relevant aspects of the dual superconducting phase.
In the next section, we will obtain the corresponding numerical solution to equations (\ref{metricansatz}) to \eqref{Aa}, subject, of course, to very specific boundary conditions. As we have already explained, this geometry is dual to a $3+1$ strongly coupled CFT in the presence of a strong magnetic field.

As our next step, we will devote the rest of this section to propose an ansatz for the complex scalar field which, as explained, is dual to a charge condensate with a non-trivial phase structure.

Let us begin by noticing that equation (\ref{psi}) is invariant under the gauge transformations
\begin{align}
\Psi \rightarrow e^{i \xi \Gamma} \Psi, & \qquad A_{\mu} \rightarrow A_{\mu} + \partial_{\mu} \Gamma, 
\end{align}
where $\Gamma$ is a differentiable function that can depend on the $x$ and $y$ coordinates.
We can choose a particular gauge so that the components of $ A_{\mu} $ no longer depend explicitly on the $y$ coordinate (of course, this is a gauge-dependent result). Then, consistently with the symmetries of our system, the ansatz for the scalar field can be written as

\begin{equation}
\Psi = R(r) F(x,y) e^{i k_z z} \, .
\label{ansatzpsi}
\end{equation}
Substituting (\ref{ansatzpsi}) into \eqref{psi}, we obtain
\begin{align}
0 = & \left[ 2 (\xi A_t + (\xi A_z-k_z) C)^2 V W - 
 2 U (V ((k_z - \xi A_z)^2 + m^2 W)) \right] F(x,y) R \nonumber \\
 & \qquad + 2 \left[  U(2 U W V' + 
   V (2 W U' + U W')) R' +  U^2 V W  R'' \right] F(x,y)  \nonumber \\
 & \qquad  + 2W U \left[- b^2 \xi^2 x^2 F(x,y) + 2 i b \xi x \frac{\partial F(x,y)}{\partial y} + \frac{\partial^2 F(x,y)}{\partial x^2}+ \frac{\partial^2 F(x,y)}{\partial y^2} \right] R,
\label{psicompleta1}
\end{align}
where ``$\prime$'' stands for $\frac{\partial}{\partial r}$. We see that (\ref{psicompleta1}) can be separated into an equation for $R(r)$ given by,
\begin{align}
-2\varepsilon U W R & =  2 \left[  U(2 U W V' + 
   V (2 W U' + U W')) R'\right] \nonumber \\ 
 & \qquad + \left[ 2 (\xi A_t + (\xi A_z-k_z) C)^2 V W \right] R \nonumber \\
 & \qquad  -\left[ 
 2 U (V ((k_z - \xi A_z)^2 - m^2 W)) \right] R \nonumber \\
 & \qquad + 2 U^2 V W  R'' , 
\label{ecr1}
\end{align}
and another for the function $F(x,y)$ that reads
\begin{equation}
\frac{\partial^2 F(x,y)}{\partial x^2} + \frac{\partial^2 F(x,y)}{\partial y^2} - b^2 \xi^2 x^2 F(x,y) + 2 i b \xi x \frac{\partial F(x,y)}{\partial y}  =  \varepsilon F(x,y),
\label{ecF1}
\end{equation}
where $\varepsilon$ is the separation constant.  To further simplify equation (\ref{ecF1}), $F(x,y)$ can be rewritten as
\begin{equation}
F(x,y) = \tilde{F}(x,y)e^{i b \xi x y},
\label{fnew}
\end{equation}
which substitution into (\ref{ecF1}) leads to
\begin{align}
\varepsilon \tilde{F}(x,y) = & \frac{\partial^2 \tilde{F}(x,y)}{\partial x^2} + \frac{\partial^2 \tilde{F}(x,y)}{\partial y^2} - b^2 \xi^2 \left(x^2  + y^2 \right) \tilde{F}(x,y)  \nonumber \\
 & \qquad + 2 i b \xi \left( x \frac{\partial \tilde{F}(x,y)}{\partial y} - y \frac{\partial \tilde{F}(x,y)}{\partial x} \right).
\label{eqx}
\end{align} 
It is worth noting that (\ref{eqx}) has the form of an equation that describes two coupled quantum harmonic oscillators with mass $m_{h} = 1/8$ and frequency $\omega = 4 b \xi$. This result has already appeared in \cite{albash2008holographic,Dector:2013dia, Pal:2019yff}, and is anticipated in the context of holographic superconductors.  The corresponding CFT interpretation, which we will review below, is well established and understood (see, for example, \cite{Hartnoll:2008kx, nakano2008critical, Natsuume:2022kic}).

In the following section, we will present and explain, in great detail, the numerical solution that describes the backreacted geometry dual to a holographic superconductor in the presence of a strong magnetic field. To understand and characterize the superconducting phase, we will also obtain a numerical solution for the complex scalar field.

\section{The Holographic Superconductor: Numerical Results}
\label{secnumresults}

In this section, we describe the details of our numerical methods and present the main results our setup, i.e., a holographic superconductor in four dimensions subjected to a strong magnetic field. In essence, the objective is to numerically solve the equations of motion derived from the holographic setup to explore the behavior of the superconducting condensate and the corresponding phase transition.

As customary in similar holographic setups, we will use the so-called ``shooting method'' as our main numerical computational technique. This is a well-known numerical method commonly used to solve boundary value problems. Loosely speaking, it consists of converting the boundary value problem into an initial value problem. Importantly, this method usually relies on the use of some analytical data to obtain the complete numerical solution.  

We see that, as stated in \cite{DHoker:2009mmn, Leo1}, the ansatz given by (\ref{metricansatz}) to (\ref{Aa}), includes as particular cases three analytical solutions: i) If $A_{\mu}=\Psi=0$, the corresponding solution is the Schwarzschild-$AdS_5$ black hole, ii) when $A_{t}\ne 0$ and $\Psi=0$, the solution describes the Reissner-Nordstrom-$AdS_5$ black hole and finally, iii) if $\Psi=0$ and  $V(r)=\frac{b}{3}$ (i.e. $F_{ij}\ne 0$), the black hole solution is the geometry given by BTZ $\times R^2$. Below, we will show how to construct a family of numerical solutions that will smoothly interpolate among these black holes, allowing us to verify that our results correctly approach those that can be obtained using such analytical backgrounds.

Since the equations of motion are singular at the horizon, identified as the locus where $U(r_h)=0$, they need to be solved by a series expansion around $r_h$. Evaluating the resulting series near the horizon will provide the necessary information to find the desired numerical solution. With this in mind, and extending the work presented in \cite{Leo1}, we introduce the expansions

\begin{equation}
\begin{split}
\label{Horizon_Series}
U(r) & = r_h (6-q^2) (r-r_h) + \sum_{i=2}^{\infty} U_i (r-r_h)^i, \\
V(r) & = V_0 + \sum_{i=1}^{\infty} V_i (r-r_h)^i, \\
W(r) & = 3 r_h^2 + \sum_{i=1}^{\infty} W_i (r-r_h)^i, \\
C(r) & = \sum_{i=1}^{\infty} C_i (r-r_h)^i,\\
P(r) & = \sum_{i=1}^{\infty} P_i (r-r_h)^i, \\
\mathcal{E}(r) & = q + \sum_{i=1}^{\infty} \mathcal{E}_i (r-r_h)^i,
\end{split}
\end{equation}
where at this point, $V_0$, $U_i$, $V_i$, $W_i$, $C_i$, $P_i$ and $\mathcal{E}_i$ are constants yet to be determined. To do this, we begin by substituting (\ref{Horizon_Series}) into equation (\ref{ecr1}). The corresponding on-shell equations, supplemented by the condition that the Hawking temperature is given by $T_h=2 \pi U'(r_h)=r_h(6-q^2)$, allow us to write any undetermined coefficients of (\ref{Horizon_Series}) in terms of the rescaled magnitude of the magnetic field $b/V_0$, the location of the horizon $r_h$, and the electric charge $q$ \cite{Leo1}. 
Moreover, with the help of the local symmetries given by (\ref{localsymmetry}), we can set $P(r_h) = C(r_h) = 0$ and guarantee that we are working in a static framework (i.e. every off-diagonal term should be zero at the horizon). This result greatly simplifies (\ref{Horizon_Series}), making our numerical computations much simpler.  

It is important to emphasize that, given our particular setup, (\ref{Horizon_Series}) can only depend on three parameters: i) the location of the black hole horizon $r_h$, ii) the electric charge $q$ and iii) the magnitude $b/V_0$ is related to the bulk magnetic field \cite{albash2008holographic}. 
Now, as mentioned above, by varying the values of these parameters, we can interpolate between the analytical black hole solutions presented in \cite{DHoker:2009mmn}, and generate a family of well-defined boundary conditions at the horizon described by (\ref{Horizon_Series}) with all the corresponding coefficients fixed. 

Our next step is to find the necessary conditions for the metric functions $U(r)$, $V(r)$, and $W(r)$, which guarantee that the desired numerical solution is asymptotically $AdS_5$. Starting from (\ref{ecr1}) and studying its near boundary behavior (i.e. $r\rightarrow \infty$), it can be shown that the metric functions take the form
\begin{equation}
\begin{split}
\label{asymp_AdS}
U(r\rightarrow \infty) &= r^2 + O[r^{-1}], \\
V(r\rightarrow \infty)& = v r^2 + O[r^{-1}], \\
W(r\rightarrow \infty)&= w r^2 + O[r^{-1}],
\end{split}
\end{equation}
where $v$ and $w$ are constants to be determined. It is easy to show that for any given value of $v$ and $w$,   (\ref{asymp_AdS}) does not satisfy the standard Brown-Henneaux boundary conditions \cite{Brown:1986nw}. However, setting $b \to \frac{b}{v}$, and with the help of (\ref{localsymmetry}), it is possible to perform a boost in the $r$ coordinate and reescale the metric functions $V(r)$ and $W(r)$ and obtain the desired asymptotic $AdS_5$ boundary behavior \cite{Leo1}. The rescaled metric functions are given by,    
\begin{align}\label{Ad5behavior}
V(r) \to \frac{V(r)}{v}\ , \qquad W(r) \to \frac{W(r)}{w}\ , 
\end{align}
where as expected, $v$ and $w$, depend only on the parameters $b/V_0$, $q$ and $r_h$. By taking advantage of this, we can perform our computations for any non-vanishing numerical value of $V_0$ and then rescale the solution appropriately. For simplicity, we take $V_0=1$. Also, notice that $B = \frac{b}{v}$ is the value of the magnetic field in the CFT.

Before moving forward, let us briefly summarize the above results. We have detailed the boundary conditions essential for determining the numerical solution for the backreacted metric. At the horizon, these conditions ensure regularity and non-singularity, facilitating the initialization of the integration process. We expanded the metric functions and gauge field components in series around the location of the horizon, $r_h$, and expressed them in terms of the relevant parameters, i.e. the magnitude of the magnetic field $b$, the electric charge $q$ and of course, $r_h$. At the asymptotic boundary, we imposed conditions that match the expected $\text{AdS}_5$ behavior \cite{Brown:1986nw}, adjusting the metric functions to ensure the correct scaling. With these boundary conditions established, we have all the necessary ingredients to find the desired solution for the backreacted metric.

In Section \ref{secbulkequations}, starting from the ansatz (\ref{ansatzpsi}), we obtained (\ref{ecr1}) and (\ref{ecF1}), which given our particular setup, describe the dynamics of the scalar field $\Psi$ propagating on the metric presented in the previous paragraphs. In what follows, we will find the corresponding numerical solution for $\Psi$ which, as explained before, in the context of the AdS/CFT correspondence, will help us to characterize the superconducting phase of the holographic superconductor under consideration.

Let us begin by studying the near boundary behavior of (\ref{ecr1}). In this particular set of coordinates, when $r\rightarrow\infty$, equation (\ref{ecr1}) reads 
\begin{equation}
m^2 R(r) - 5 r R'(r) - r^2 R''(r) = 0.
\end{equation}
The corresponding solution can be written as,
\begin{equation}
R(r \rightarrow \infty) = \frac{\psi_{+}}{r^{\Delta_{+}}} +\frac{\psi_{-}}{r^{\Delta{-}}} +\cdots +,
\label{fronteraR}
\end{equation}
where $\psi_{+}$ and $\psi_{-}$ are constants yet to be determined, and the exponents $\Delta_{\pm}$ are given by, 
\begin{equation}
\Delta_{\pm} =2 \left( 1 \pm \sqrt{1+ \frac{m^2}{4}} \right).
\label{delta}
\end{equation}  
When $0 > m^2 \ge -3$ , $\Psi_+$ and $\Psi_-$ describe normalizable modes. Moreover, according to the holographic correspondence, from the $\Psi_+$ mode, we can read off the expectation value of the operator dual to $\langle\mathcal{O}\rangle$ whilst the $\Psi_-$ mode is dual to a source for the operator $\mathcal{O}$ \cite{Hartnoll:2008kx,Hartnoll:2008vx,Herzog:2009xv,Hartnoll:2009sz}. To carry on with our computation, we choose $m^2 = -3$; firstly, because it greatly simplifies \eqref{fronteraR}, and secondly, as explained in \cite{albash2008holographic}, because the general qualitative behavior of our numerical solution does not depend on this specific value. 
Then, setting $m^2 = -3$, equation \eqref{fronteraR} reduces to,
\begin{equation}
R(r \rightarrow \infty) = \frac{\psi_-}{r} + \frac{\psi_+}{r^3}\ .
\label{Rinf}
\end{equation}
Choosing $\psi_- = 0$ as boundary condition guarantees, from the dual field theory point of view, that the condensate $\langle\mathcal{O}\rangle$ can be turned on without being sourced.

Now, to determine the necessary boundary condition at the horizon, we follow a similar procedure. That is, starting from the on-shell equation of motion \eqref{ecr1}, supplemented by the condition $T_h=r_h(6-q^2)$, we obtain,  
 
\begin{equation}
R(r \rightarrow r_h) = R_0 + R_1 (m^2 + b \xi)(r -r_h) + \sum_{i=2}^{\infty} R_i (r-r_h)^i,
\label{solR}
\end{equation}
where $R_0$, $R_1$, and $R_i$ are constants to be determined, and $m$, $\xi$ are the mass and charge of the complex scalar field respectively. It can be shown that the coefficients $R_1$ and $R_i$ are proportional to $R_0$ and can be written in terms of the parameters $r_h$, $b$ and $q$; this implies that there are no undetermined coefficients in (\ref{solR}). It is worth mentioning that, similarly to the role played by $V_0$ in (\ref{Horizon_Series}), the qualitative behavior of (\ref{solR}), does not depend on the value of $R_0$. The results that we will present below were obtained using $R_0=1$.

Before we begin to present our main results, some comments are in order: i) Given (\ref{Horizon_Series}) and (\ref{solR}), i.e., the expansion at the horizon for the metric, the Maxwell field, and the scalar field, it is tempting to assume that the corresponding numerical solutions will depend on the parameters present on the equations just mentioned, that is, $r_h$, $q$ and $b$. It was only after a better understanding of our system,  supported by numerical data, that we were able to conclude that only two of these parameters were independent. For reasons that will be clarified below, we have chosen to work with $b\equiv b(r_h)$ and $q\equiv q(r_h)$.  ii) The approximation $\phi \ll 1$ is equivalent to taking a large $\text{AdS}_5$ radius compared to $r_h$, which in turn, means that the temperature of our system is well below the critical temperature at which the condensate vanishes. This restricts our setup so that we can only explore the behavior of a superconductor at low temperatures. 

To obtain solutions satisfying $\psi_- = 0$, we first set a value for $b$ and then search for a corresponding value of $q$ that yields $\psi_- = 0$. In this form, for each value of $b$, there is only one corresponding value of $q$ that allows us to satisfy $\psi_- = 0$. Thus, we are effectively left with only one free parameter.  Besides, if we set $b$ above a certain value, $\psi_-$ no longer vanishes for any value of $q$, thus defining a critical value $b_c$.

Once we have obtained values for $ b $ and $ q(b) $ for which $ R (r) $ is normalizable and satisfies the desired boundary conditions, we can evaluate all the relevant physical quantities.

\begin{figure}
\includegraphics[width=0.6\textwidth]{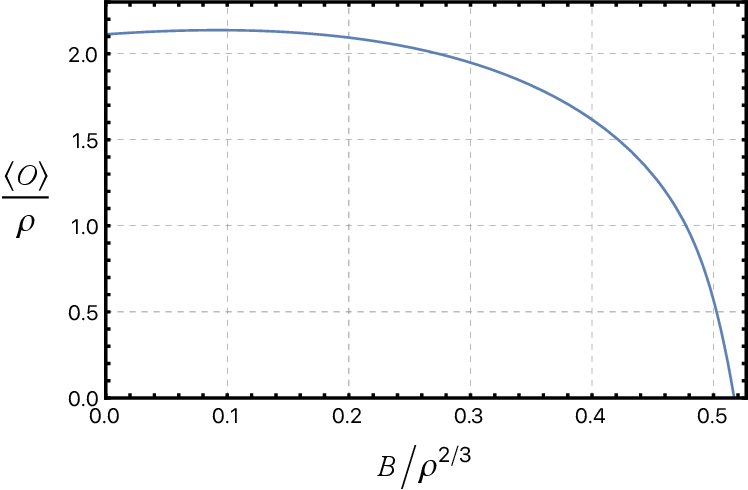}
\caption{Condensate $\langle \mathcal{O}\rangle$ vs. magnetic field $B$, both normalized by the charge density $\rho$.  With $k=1$ and $\xi = 5$.}
\label{BPsi_rho}
\end{figure}

Figure \ref{BPsi_rho} shows the condensate as a function of the magnetic field $B$, both normalized by the charge density $\rho$, so that the results are presented in terms of dimensionless quantities. As the magnetic field $B$ increases, the condensate decreases until it vanishes at a critical magnetic field $B_c \approx 0.52 \rho^{2/3}$. For $B > B_c$, the only viable solution that is consistent with the background described by \eqref{metricansatz} and \eqref{Fansatz} and satisfies the boundary conditions is $\Psi=0$. This establishes an upper limit for the magnetic field that a superconductor can sustain.

\begin{figure}
\includegraphics[width=0.6\textwidth]{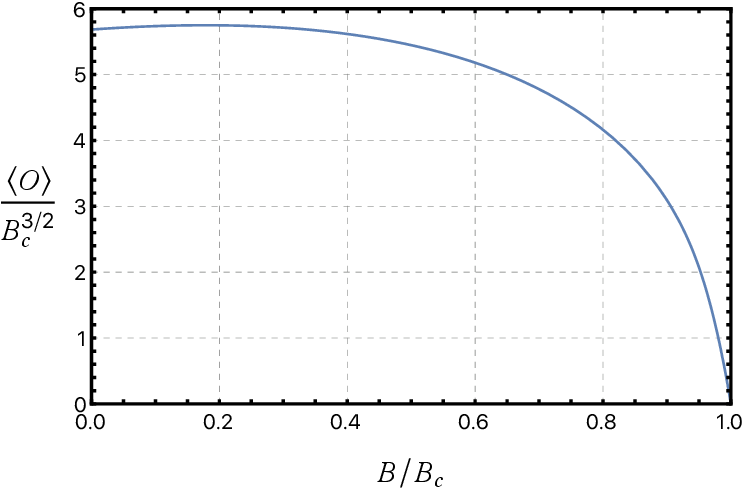}
\caption{ Condensate $\langle \mathcal{O}\rangle$ vs. magnetic field  $B$ both normalized by the critical magnetic field $B_c$. With $k=1$ and $\xi = 5$.}
\label{BPsi}
\end{figure}

Once the critical magnetic field $B_c$ has been determined, it can be used as a normalization parameter. Figure \ref{BPsi} depicts the condensate as a function of the magnetic field intensity, with both quantities normalized by $B_c$. The resulting graph exhibits a similar pattern to the conventional plots of condensate versus temperature in holographic superconductors \cite{Horowitz:2008bn}.

\begin{figure}
\includegraphics[width=0.6\textwidth]{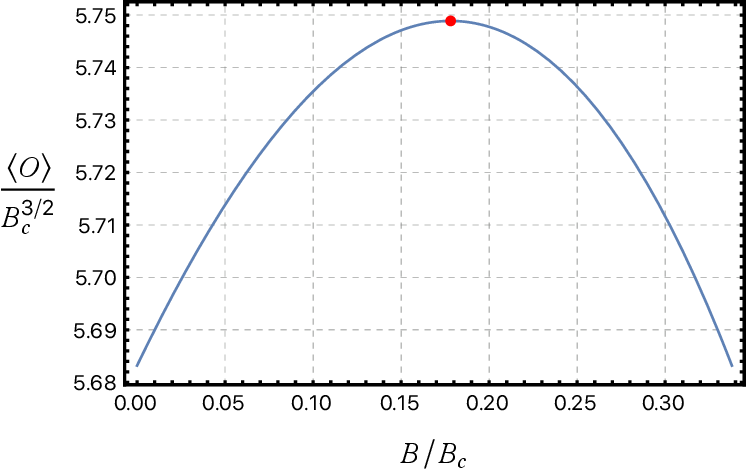}
\caption{ Condensate $\langle \mathcal{O}\rangle$ vs. magnetic field $B$ with $k=1$ and $\xi = 5$ in the region where $B/B_c $ is small.}
\label{BPsi2}
\end{figure}

In the region where $B \ll B_c$, as shown in Figure \ref{BPsi2}, we observe that for $B$ in the range $[0, \sim 0.178)$, the condensate increases and reaches a global maximum. This behavior can be interpreted as the superconducting phase becoming more difficult to break. Subsequently, the condensate decreases for bigger values of $B/B_c$. We speculate that this favorable behavior for the condensate formation in the region of $B/B_c \sim 0$ is related to the magnetic catalysis phenomenon mentioned in \cite{Bolognesi:2011un}.

The complete solution for the scalar field $\Psi$ includes the $x$ and $y$ dependent part, which can be written as the product of two functions $F_0(x,y) = X(x)Y(y)$.  Each of these functions takes the form of a Gaussian distribution, as described by \eqref{eqx}. This suggests that the spatial profile of the scalar field has a Gaussian shape in both the $x$ and $y$ directions, which after setting $b \to \frac{b}{V_0} = B$, take the form
\begin{align}
X(x) = \left( \frac{B \xi}{2 \pi} \right)^{\frac{1}{4}}  e^{-\frac{B \xi x^2}{4}}, && Y(y) = \left( \frac{B \xi}{2 \pi} \right)^{\frac{1}{4}} e^{-\frac{B \xi y^2}{4}},
\end{align}
such that the value of the condensate in the field theory considering its spatial dependence is
\begin{equation}
\langle \mathcal{O}_X \rangle = \psi_{+} \exp(-\frac{x^2}{2 \sigma^2}) \exp(-\frac{y^2}{2 \sigma^2}),
\end{equation}
with standard deviation $\sigma = \frac{\sqrt{2}}{\sqrt{B \xi}}$.

When the magnetic field $B$ is very small compared to the critical magnetic field $B_c$, the condensate has negligible dependence on $x$ or $y$ and is essentially constant. However, as the magnetic field grows, the condensate becomes localized in the $x$ and $y$ directions. This behavior can also be expressed in terms of the radial coordinate of the field theory $X= \sqrt{x^2 + y^2}$.

Although translational invariance was initially imposed on the metric and the gauge field in the $x, y$ directions, the condensate exhibits a spatial dependence characterized by a Gaussian profile. The standard deviation represents a length scale that determines the extent to which the condensate becomes significantly small with respect to the charge density. In the limit of vanishing magnetic field, $B \rightarrow 0$, the standard deviation $\sigma$ tends to infinity. This indicates that in the absence of a magnetic field, the condensate occupies the entire space. On the other hand, as the magnetic field strength approaches infinity, $B \rightarrow \infty$, $\sigma$ approaches zero, implying that the condensate becomes highly localized. This behavior is reminiscent of the Meissner effect, a characteristic phenomenon observed in superconductors \cite{London1935TheEE}.

\begin{figure}
\includegraphics[scale=0.8]{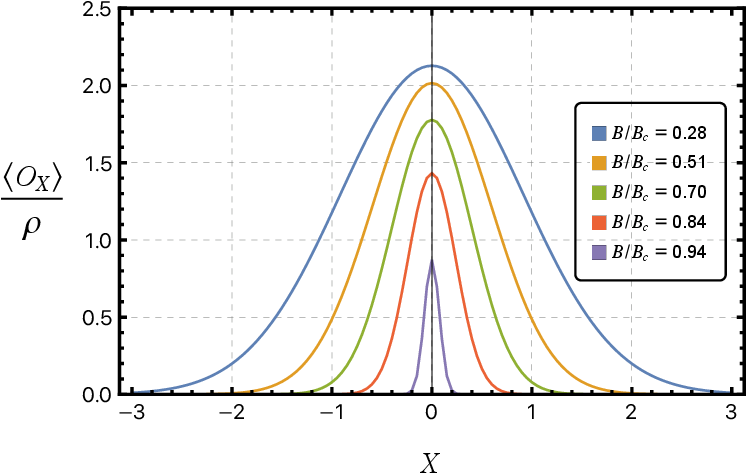}
\caption{Spatial profile for the condensate $\langle \mathcal{O}_X \rangle$ for different values of $ B/B_c $.}
\label{equis}
\end{figure}

\section{Results and Final Remarks}
\label{secresults}

In this section, we summarize the key findings from our numerical analysis of 4-dimensional holographic superconductors in the presence of a strong magnetic field.  

Our results demonstrate a clear relationship between the strength of the magnetic field and the behavior of the superconducting condensate. Except in the region $B\ll B_c$, where the condensate increases with the magnetic field and we speculate a connection with the phenomenon of magnetic catalysis (e.g. \cite{Bolognesi:2011un, shovkovy2013magnetic}), as the magnetic field $B$ increases, the condensate $\langle \mathcal{O} \rangle$ diminishes, eventually vanishing at a critical magnetic field $B_c$. This critical field marks the upper limit for superconductivity in our model. The dependence of the condensate on the magnetic field is depicted in Figure \ref{BPsi} of our study, showing a smooth transition to the normal phase as the field strength increases.

We observed that the spatial profile of the condensate in the dual field theory follows a Gaussian distribution. The standard deviation of this distribution decreases with increasing magnetic field, implying that the condensate becomes more localized. In the absence of a magnetic field, the condensate is uniformly spread, whereas at higher fields, it shrinks significantly, suggesting a behavior akin to the Meissner effect where the magnetic field penetrates the superconductor and suppresses superconductivity.

Based on the analysis conducted of the physical quantities studied, it is inferred that  the critical temperature $T_c$ decreases with increasing magnetic field strength. This is consistent with the general behavior of type II superconductors, where higher magnetic fields can suppress the superconducting state. The numerical solutions obtained suggest that for fields exceeding the critical value $B_c$, the system transitions entirely to the normal phase, indicating the breakdown of superconductivity under strong magnetic influences.

While we considered the full backreaction of the Maxwell field, the charged scalar field was treated perturbatively. This approach allowed us to focus on the essential physical effects without the computational complexity of a fully backreacted scalar field. Our results, therefore, primarily capture the leading-order effects of the scalar perturbation on the background geometry and gauge field.

Our investigation into the holographic superconductors in a four-dimensional spacetime with a strong magnetic field has yielded significant insights into the interplay between superconductivity and magnetic fields in strongly coupled systems. The use of the holographic correspondence has proven to be a powerful tool in modeling these complex phenomena, providing a window into the behavior of high-temperature superconductors and other strongly correlated materials.

The findings from this study highlight the potential of holographic methods to uncover new aspects of superconductivity that are difficult to access through traditional condensed matter approaches. Future work could extend this analysis by considering the full backreaction of the scalar field and exploring other parameter regimes, such as varying the charge of the scalar field or introducing additional fields to mimic more complex interactions.

\section{Acknowledgements}

The work of M.C. and M.R. is partially supported by Mexico's National Council of Humanities, Sciences, and Technologies (CONAHCyT) grant A1-S-22886, and DGAPA-UNAM grant IN116823. The work of C.D. and M.R. is supported by CONAHCyT Ph.D. fellowships.

\newpage
\bibliography{biblio.bib}
\bibliographystyle{unsrt}
\end{document}